\documentclass[10pt]{article}
\usepackage{latexsym,graphicx}
\usepackage{latexsym,graphicx}
\newcommand{\be}{\begin{equation}}
\newcommand{\ee}{\end{equation}}
\def\n{\noindent}
\catcode `\@=11 \catcode `\@=12
\begin{document}
\begin{center}
\large{\bf {Is Hubble's Expansion due to Dark Energy}}\\
\vspace{10mm}
\normalsize{R. C. Gupta $^1$ and Anirudh Pradhan $^2$} \\
\vspace{5mm} \normalsize{$^1$ GLA University, Mathura-281 406, India} \\
\normalsize{E-mail: rcg\_iet@hotmail.com}\\
\vspace{5mm} \normalsize{$^2$ Department of Mathematics, Hindu
P. G. College, Zamania-232 331, Ghazipur, U. P., India} \\
\normalsize{E-mail: pradhan@iucaa.ernet.in}\\
\end{center}
\vspace{10mm}
\begin{abstract}
{\it The universe is expanding}  is known (through Galaxy observations) since 1929 through Hubble's discovery 
($V = H D$). Recently in 1999, it is found (through Supernovae observations) that  the universe is not simply 
expanding but is accelerating too. We, however, hardly know only $4\%$ of the universe. The Wilkinson Microwave 
Anisotropy Probe (WMAP) satellite observational data suggest $73\%$ content of the universe in the form of 
dark-energy, $23\%$ in the form of non-baryonic dark-matter and the rest $4\%$ in the form of the usual baryonic 
matter. The acceleration of  the universe is ascribed to this dark-energy with bizarre properties (repulsive-gravity). 
The question is that whether Hubble's expansion is just due to the shock of big-bang \& inflation or it is due to 
the repulsive-gravity of dark-energy? Now, it is believed to be due to dark-energy, say, by re-introducing the 
once-discarded cosmological-constant $\Lambda$. In the present paper, it is shown that `the formula for acceleration 
due to dark-energy' is (almost) exactly of same-form as `the acceleration formula from the Hubble's law'. Hence, it 
is concluded that: yes, `indeed it is the dark-energy responsible for the Hubble’s expansion too, in-addition to the 
current on-going acceleration of the universe'.
\end{abstract}
\smallskip
\n Key words: Cosmology, Dark Energy, Hubble's expansion \\\
\n PACS: 98.80.-k, 95.36.+x, 04.20.-q\\
\section{Introduction}
The scenario before the discovery, in early 20th century (in 1929), of Hubble's expansion was almost stagnant 
\& static in astrophysics \& cosmology. Though revolutionary changes occurred when Ptolemy geo-centric model was 
replaced by Copernicus helio-centric model, and advancement made then-after with the work of  many scientists 
(such as Galileo, Kepler, Newton \& Einstein) too are notable; but as a whole the research situation was progressing 
slowly, it was almost stagnant. Even the universe was considered, for sure, to be static. Einstein theory, through 
Friedman's equation, was predicting the expanding (or contracting) universe; but Einstein himself, being 
over-confident as per the then prevailing belief that the universe is static, introduced a fuzz-factor 
(cosmological constant $\Lambda$) into his equation to make the universe static. Einstein, later, after hearing the 
Hubble's expansion, regretted the introduction of $\Lambda$ and said it to be his `greatest blunder' of life 
\cite{ref1,ref2}. Though Einstein abandoned $\Lambda$ but some other scientist kept it alive, however, for one 
reason or the other. The expansion implies that in the past the universe was smaller. If we rewind the conceptual-film 
of expansion, we reach to a point (singularity) wherein the so-called big-bang occurred. If we run the film forward 
from the big-bang, after about $1$ lac years we pass through an era of de-coupling (of matter \& radiation), the relic 
of which is known as cosmic-microwave-background (CMB) radiation and it is considered as the firm evidence of  big-bang
theory [1-3]. Soon after the big-bang, within extremely-short-time there was a super-rapid exponential-expansion, 
called inflation [1-3]. Hubble's expansion is quite different, however. Much later at the end of 20th century 
(in 1999), with the Supernovae observations, scientists came to conclusion that universe is not only expanding but 
accelerating too [4-9]. These observations of type Ia supernovae (SNe Ia) suggest that the accelerated expansion of 
the universe is possibly powered by a smooth energy component with negative pressure dubbed as dark energy. Therefore, 
the cosmological constant $\Lambda$ re-appeared once again, as vacuum-energy as a candidate for dark-energy causing 
the acceleration. Note that the Hubble's expansion was reported in 1929, 70 years earlier than the recognition of 
dark-energy in 1999 responsible for the current-ongoing acceleration of  the expanding-universe. Is the dark-energy 
also responsible for Hubble's expansion, in addition to the acceleration of the universe? Moreover, what is the 
content and constituent(s) of the universe  and that what is that which expands in the expanding-universe? \\

It is rather unbelievable but true, that in general, we even don't  know that `we don't  know with what (constituents) 
the universe is made of'. It is surprising that we hardly know only $4\%$ of the universe. Rest of the universe is 
said to be made up of $73\%$ of dark-energy and $23\%$ of dark-matter. Though science (including astrophysics \& 
cosmology) has progressed a lot; but even the scientists don't know exactly what are the dark-energy \& dark-matter, 
albeit a few theories have been suggested in literature. Dark-energy [4-9] is responsible for acceleration of the 
expanding universe; whereas dark-matter [10-14] is said to be necessary as extra-mass of bizarre-properties to explain 
the anomalous rotational-velocity of galaxy. \\
\begin{figure}[ht]
\centering
\includegraphics[width=15cm,height=10cm,angle=-0]{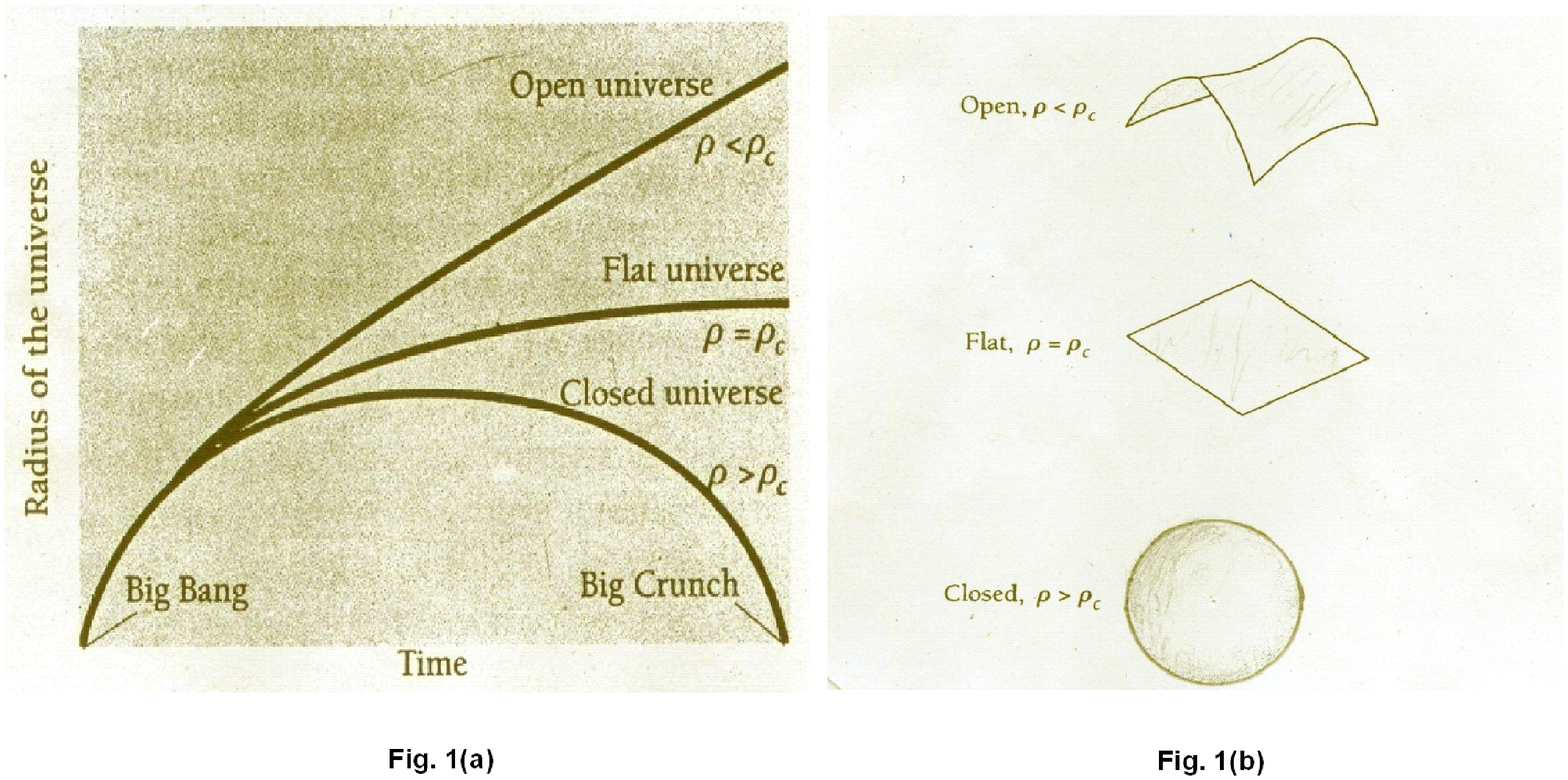} \\
\caption{(a) Fate of Closed, Flat and Open Universe (b) The 2-dimensional Analogue of the Universe 
(sphere, flat-plate, horse-saddle)}
\end{figure}
\section{About the Content and Constituent(s) of the Universe}
How much matter and energy are there in the universe ? It is now well established that universe-expansion began with 
a big-bang [1-3]. The ultimate fate of the universe (Fig.1) depends [1-3] on the 
universe's matter \& energy density ($\rho$) as compared to a certain value called critical-density ($\rho_{c}$). 
If $\rho > \rho_{c}$, the universe is said to be `closed'; and its expansion will slow down (decelerate) and start 
contracting leading finally to a big-crunch (meaning hot-death of the universe). If $\rho < \rho_{c}$, the universe is 
said to be `open'; and will expand forever even much faster (leading the universe to cold-death). If $\rho = \rho_{c}$, 
the universe is said to be `flat'; and will continue to expand but not that-fast to lead to cold-death soon. 
The density-ratio (Omega) $\Omega = \rho/\rho_{c}$, determines (the nature \& fate of the universe) that whether the 
universe is closed ($\Omega > 1$), open ($\Omega < 1$) or flat ($\Omega = 1$). It has been estimated that the universe 
would have collapsed (to hot-death) much sooner than the present-age of the universe if  $\Omega > 1$; and it would 
have cooled down (to cold-death) much earlier than the present-age of the universe if $\Omega < 1$. The present-age 
(14 billion years) constraint of the universe, compel the scientists to believe that $\Omega = 1$, i.e., the universe 
must be flat [1-3]. \\

Once agreed-upon that the universe density $\rho = \rho_{c}$, the next question arises that `what is the universe 
made of'? Estimation of visible-type matter like galaxies, stars, planets etc. hardly leads only to about 
$2\%$ of $\rho_{c}$; and when other all such things like inter-galactic gases, black-hole, white-dwarf, neutron-stars 
etc. are also included, the estimate hardly reaches a mere $4\%$ of $\rho_{c}$. What is then $96\%$ of the 
remaining-part? It seems invisible and unknown, hence thought as dark constituent(s). Scientists have, presently, 
estimated that the major-chunk of the universe is repulsive-gravity type dark-energy (about $73\%$) causing the 
universe's  accelerated expansion [4-9], and the rest is non-baryonic invisible but gravitating 
dark-matter (about $23\%$) causing anomalous high rotational-speed of galaxies [10-14]. \\

It is intriguing that most abundant stuff (dark energy) is least understood. The major candidate for the dark-energy 
is considered to be the cosmological constant $\Lambda$ (vacuum-energy), though other theories  have been proposed, 
for dark-energy having wide-range origin(s) ranging from tiny-nucleus \cite{ref15} to mighty brane-world \cite{ref16}.

\section{Acceleration due to dark energy}
Initially it was thought that the universe would be decelerating due to gravity inside, but now it is well established 
from several clues [4-9] such as Supernovae observations that the universe is actually accelerating 
due to repulsive gravity of dark-energy. The deceleration-parameter $q$ is defined as follows in Eq. (\ref{eq1}). 
(Note that even though universe is actually accelerating, but the old name deceleration-parameter retained; but $q$ 
comes out to be actually negative, implying that $\ddot{S}$ is positive i.e., universe is accelerating). Note that 
though Scale-factor S (ratio of the co-moving distance at previous-time at $Z > 0$ to the co-moving distance at 
present-time $Z = 0$) is dimensionless whereas Size-of-universe (or co-moving- distance) has dimension of  
length; but sometimes all these are denoted by the same symbol S, mainly in view that the scale-factor is 
proportional to the size-of-universe (co-moving-distance) and incidentally both the words begin with the letter S. 
Sometimes, as in references \cite{ref15,ref17}, scale-factor and universe-size are denoted by symbol `a' (but we 
can not use such symbol here because in this paper `a' is used for acceleration). So it is better for clarity, if 
all these are denoted by different symbols as follows; scale-factor as S, co-moving distance between two points as D, 
universe size as $D_{max}$, all being function of time due to expansion of universe. Thus scale factor $S(t) = 
D(t) / D_{0} $ or simply $S = D / D_{0}$ where D is the co-moving distance in the past (at $Z > 0$) and $D_{0}$ is 
the co-moving distance at present (at $Z = 0$). Hence it is obvious that S is proportional to D. Universe-size is 
$D_{max}$ i.e., the distance of the visible universe-horizon, such that as per Hubble's law $V_{max} = c = H D_{max}$. 
(Note that even if universe-tip may be moving with speed higher than light-speed, as it was during inflation, the 
observable `visible' horizon will be limited by the equation $c = H D_{max}$). \\\\
The deceleration parameter q is conventionally defined by 
\begin{equation}
\label{eq1} q = - \frac{S \ddot{S}}{\dot{S}^{2}}.
\end{equation} 
Putting the experimental (See the Ref. \cite{ref18}) value of $q = −0.67$, the expression for the acceleration 
($a_{d}$) of the universe (or galaxy as the case may be) due to dark-energy is given by,
\begin{equation}
\label{eq2} a_{d} = \ddot{S} = 0.67\left(\frac{\dot{S}^{2}}{S}\right).
\end{equation}  
But in Eq. (\ref{eq2}) what should we use for S ? It seems for galaxy observations, more appropriately S should be 
(being proportional) the co-moving-distance D, say, between the observed galaxy (say, Andromeda galaxy) from the 
earth (situated in the Milky-way galaxy). Hence, the galactic acceleration due to dark energy can be obtained as 
follows by replacing S by D in the previous equation,
\begin{equation}
\label{eq3} a_{d} = \ddot{D} = 0.67\left(\frac{\dot{D}^{2}}{D}\right).
\end{equation}
\section{Acceleration due to Hubble's expansion}
Note that from Hubble's law of the expansion, velocity is proportional to distance  i.e., $V = H D$, H  being 
Hubble's constant. Hubble’s law $V = H D$ is rewritten as $\dot{D} = H D$. This also gives the acceleration 
\begin{equation}
\label{eq4} \ddot{D} = H \dot{D} = \left(\frac{\dot{D}}{D}\dot{D}\right) = \frac{\dot{D}^{2}}{D},
\end{equation}
which is (almost) exactly of the same-form (equivalent) as the Eq. (\ref{eq3}). Thus, this also reinforces the 
understanding that the co-moving distance D is proportional to scale-factor or vice-verse. This also indicates 
an important possibility that the Hubble’s expansion is due to dark-energy. Therefore, galactic acceleration due 
to Hubble's expansion  is,
\begin{equation}
\label{eq5} a_{h} = \ddot{D} = \left(\frac{\dot{D}^{2}}{D}\right).
\end{equation}
\section{Is $a_{d}$ and $a_{h}$ are same or similar?}
The two equations, one for the galactic acceleration $a_{d}$ (Eq. 3) due to dark-energy and the other for the 
galactic acceleration (Eq. 5) due to Hubble’s law, are of (exactly) the same-form, possibly it could be equal too, 
but the accelerations differ by a little factor of 0.67. The small factor 0.67 may be ignored, albeit it may due 
to some reason (coincidence). It, however, seems that the Hubble's expansion as well as its acceleration is  due 
to the dark-energy.  
\section{Is Hubble's expansion due to aftermath-effect of Big-bang or due to Dark-energy?}
The big question is that whether Hubble's expansion is just due to the shock of big-bang \& inflation or it is due 
to the repulsive-gravity of dark-energy? Earlier it was thought that the Hubble’s expansion is the aftermath effect 
since big-bang \& inflation, but now it is believed to be (accelerating) due to dark-energy. The similarity in 
Eq. (\ref{eq3}) \& Eq. (\ref{eq5}) is not a mere coincidence; it indicates that Hubble's expansion is indeed due 
to dark-energy. The expansion occurred much earlier; the acceleration, however, is comparatively a recent phenomenon 
much later (at  $Z = 0.5$) than the galaxy-formation era (at $Z = 3$), as described in reference \cite{ref15} in 
accordance with the astrophysical constraints \cite{ref19}. The common understanding of the universe-expansion is 
through re-introducing the once-discarded cosmological-constant $\Lambda$; though there are other theories/candidates 
\cite{ref20} for dark-energy such as  scalar-field, nuclear-energy, brane-world. What is that  which is expanding?  
It is the fabric of universe or the scale-factor $ S (= \frac{D}{D_{0}})$ or the co-moving distance D or its 
maximum-value $D_{max}$ (the universe) expanding, as mentioned in the section-3.
\section{The $2/3$ Coincidences}
There are several coincidences of appearance of a factor $2/3 (= 0.67 \mbox{or nearly} 70\%)$ in the physics \& cosmology 
(which may have some reasons, a few of which are explained in reference \cite{ref15}) . These $2/3$ coincidences 
are listed as follows. There seems to be some reasons for it (as just indicated), or these coincidences of $2/3$ are 
purely by chance (coincidence)?  The regime of the $2/3$ coincidences appear to be too wide, ranging from tiny 
strings \& quark  to  mighty  universe \& cosmos.
\begin{itemize}
\item
For dark-energy (quintessence) the equation of state parameter $w = - 2/3$.
\end{itemize}
\begin{itemize}
\item
Dark-energy is about $2/3$ (about $70\%$) of the universe-content.
\end{itemize}
\begin{itemize}
\item
Deceleration parameter (Experimental \cite{ref18}) $q = - 0.67$ i.e., $q = - 2/3$.
\end{itemize}
\begin{itemize}
\item
As per the Eqs. (3 \& 5) mentioned in this paper $a_{d} = 0.67 a_{h}$ or $a_{d} = \frac{2}{3}a_{h}$.  
\end{itemize}
\begin{itemize}
\item 
Hubble's age $t_{H} (= 1/H)$ and present age $t_{0}$ (for matter universe) are related as $t_{0} = (2/3)t_{H}$. 
Most likely this could be to the answer for the difference of 0.67 ( = 2/3) in the equations (3 \& 5).
\end{itemize}
\begin{itemize}
\item
Our Solar-system  is situated at about $2/3$ radial-distance from the centre of the milky-way galaxy.
\end{itemize}
\begin{itemize}
\item
Planetary-orbit R and its time-period T  are linked  with $2/3^{rd}$ power Kepler's law as $R \propto T^{2/3}$.
\end{itemize}
\begin{itemize}
\item
About $2/3^{rd}$ of the earth-surface is covered by water (sea).
\end{itemize}
\begin{itemize}
\item
Up-quark has a fractional charge of  $+ 2/3$ .
\end{itemize}
\begin{itemize}
\item
In 10-dimensional super string theory, the ratio of the usual 4-dimensional space-time to the 6 curled-up 
dimensions  is  $4/6 = 2/3$.
\end{itemize}
\section{Conclusions}
Hubble's expansion is a slow expansion of universe and from galactic-observation Hubble's famous formula is 
$ V = H D$ or $\dot{D} = H D$. This also leads to a galactic-acceleration $a_{h} = \ddot{D} = (\dot{D}^{2}/D)$. 
This incidentally is of the same-form as obtained by the formula of deceleration (acceleration) parameter; the 
acceleration caused by dark-energy is as $a_{d} = \ddot{D} = 0.67 (\dot{D}^{2}/D)$. These two equation are of 
exactly same-form, except by a factor of 2/3, the coincidence of  this yet to be fully explained. Moreover, in 
view of the striking-similarity in the equations (3 \& 5), it is concluded that: yes, `it is the dark-energy 
responsible for Hubble's expansion too, in-addition to the current on-going acceleration of the universe'.
\section*{Acknowledgements}
The authors thank Sushant Gupta for his comments \& view. The authors also thank UPTU, Lucknow, GLA University, 
Mathura and IUCAA, Pune for direct \& indirect support \& facility.

\end{document}